\documentclass[preprint,showpacs,preprintnumbers,amsmath,amssymb,aps]{revtex4}
\usepackage{amssymb}
\usepackage{epsfig}
\usepackage{psfrag}
\usepackage{graphicx}
\usepackage{color,graphics}

\begin{document}

\title{Ion-acoustic cnoidal waves in a quantum plasma }
\author{S. Mahmood$^{\ast 1,2}$ and F. Haas$^{1}$ \\
$^{1}$Physics Institute, Federal University of Rio Grande do Sul, \\
915051-970, Porto Alegre, RS, Brazil\\
$^{2}$Theoretical Physics Division (TPD), PINSTECH \\
P .O. Nilore Islamabad 44000, Pakistan}

\begin{abstract}
\noindent Nonlinear ion-acoustic cnoidal wave structures are studied in an
unmagnetized quantum plasma. Using the reductive perturbation me\-thod, a
Korteweg-de Vries equation is derived for appropriate boundary
conditions and nonlinear periodic wave solutions are obtained. The
corresponding analytical solution and numerical plots of the
ion-acoustic cnoidal waves and solitons in the phase plane are presented
using the Sagdeev pseudo-potential approach. The variations in the nonlinear
potential of the ion-acoustic cnoidal waves are studied at different values
of quantum parameter $H_{e}$ which is the ratio of electron plasmon energy
to electron Fermi energy defined for degenerate electrons. It is found that
both compressive and rarefactive ion-acoustic cnoidal wave structures are
formed depending on the value of the quantum parameter. The dependence of
the wavelength and frequency on nonlinear wave amplitude is also presented.
\end{abstract}


\maketitle

\section{Introduction}

The study of nonlinear wave propagation in quantum plasmas has gain
importance due its application in understanding the particle or energy
transport phenomenon on short scale lengths i.e., in micro and nano scale
electronic devices and in dense compact stars \cite{r2,r3,r4,r5,r6,r7,r8}.
Typically, the quantum effects in plasmas become important when the Fermi
temperature, which is related to the particles density, becomes equal or
greater than the system's thermal temperature or the inter-particle distance
becomes smaller or of the same order of the particle's de Broglie thermal
wavelength. In order to study the dynamics in quantum plasmas, the quantum
hydrodynamic (QHD) model is frequently used \cite{r2,r3,r7}. The QHD model
consists of a set of equations describing the transport of charge, momentum
and energy in a quantum charged particle system interacting through a self
consistent electromagnetic field. In QHD model, the quantum effects appear
through the quantum statistical (Fermi) pressure and the Bohm potential (due
to quantum diffraction or tunneling effects). The QHD is useful to study
collective effects on short scale lengths and has its limitation for systems
that are large compared to the Fermi Debye lengths of the species in the
system. Quantum ion-acoustic waves were investigated by Haas\textit{\ et al. 
}\cite{r2,r3} using the QHD model. They derived a Korteweg-de Vries (KdV)
equation in the weakly nonlinear amplitude wave limit for studying the
propagation of ion-acoustic solitons in a quantum plasma. It is reported
that the compressive or rarefactive soliton solution depends on a quantum
parameter ($H_{e}$) defined for degenerate electrons, which is the ratio of
the electron plasmon energy to the Fermi energy. In the fully nonlinear
regime, the existence of periodic traveling wave patterns were reported for
ion-acoustic waves in quantum plasmas. The arbitrary amplitude ion-acoustic
solitary waves using the QHD model in electron-ion quantum plasmas with the
Sagdeev potential approach has also been investigated \cite{r9}. Nonlinear
electrostatic wave structures such as solitons, envelope and shocks have
been studied in quantum electron-ion (EI) plasmas \cite{ali} but no one has
reported the propagation of cnoidal wave structures in quantum plasmas. The
purpose of the present work is to investigate the formation of ion-acoustic
cnoidal waves in a quantum EI plasma using the well known QHD model.

In classical plasmas, a lot of research work has been done in studying
nonlinear ion-acoustic wave structures such as solitons, cnoidal waves and
envelopes. Solitons are single pulse structures which are formed due to the
balance between nonlinearity and dispersion effects in the system \cite{r10}
and they consist of isolated hump or dip like wave profile with no rapid
oscillations inside the packet. Envelope structure contains both fast
and slow oscillations, obtained when nonlinearity balances the
wave group dispersion effects. The envelope is a localized modulated wave
packet whose dynamics is governed by the nonlinear Schr\"{o}dinger (NLS)
equation \cite{r11,r12}. The periodic (cnoidal) wave is the exact nonlinear
periodic wave solution of the KdV equation with appropriate boundary
conditions. These solutions of the KdV equation are also termed as cnoidal
waves because they are written in terms of Jacobian elliptic-function cn. In
general, the nonlinear periodic waves are expressed in terms of Jacobian
elliptic-functions such as dn, sn or cn, and the nonlinear dn waves are
believed to be generated in the defocusing region of the ionospheric plasmas 
\cite{r15,r16,r17,r18}. The ion-acoustic soliton and double layer structures
are observed in auroral and magnetospheric plasmas and also nonlinear
periodic wave signals appear frequently in these observations \cite{r19}.
The periodic signals are also observed at the edge of the tokamak plasma, 
which can be described by cnoidal waves \cite{r20}. Kono \textit{et al}. 
\cite{r21} studied the higher order contributions in the reductive
perturbation theory for the nonlinear ion-acoustic wave propagation under
the periodic boundary condition. The nonlinear periodic wave solution for
small amplitude Langmuir waves in electron-ion plasmas was studied by
Schamel \cite{r22}. Jovanovic and Shukla \cite{r23} presented a solution in
the form of a cnoidal wave provided the minimum value of the electrostatic
potential remain finite in studying coherent electric field structures.
Prudskikh \cite{r231} studied the ion-acoustic nonlinear periodic waves in
dusty plasmas. The ion-acoustic cnoidal wave and the associated nonlinear
ion flux in dusty plasmas was studied by Jain \textit{et al. }\cite{r232}.
They derived the coupled evolution equations for the first and second order
potentials for ion-acoustic waves in unmagnetized dusty plasmas using
reductive perturbation method with appropriate boundary conditions. Kaladze 
\textit{at al}. \cite{r24} investigated acoustic cnoidal waves and solitons
in unmagnetized pair-ion (PI) plasmas consisting of the same mass ion
species with different temperatures. They reported the formation of both
compressive and rarefactive cnoidal wave structures in PI plasmas which
depends on the temperature ratio of PI species. Recently, Kaladze and
Mahmood \cite{r25} studied the effect of positrons density and nonthermal
parameter kappa on the propagation of the ion-acoustic cnoidal waves in
electron-positron-ion plasmas. Saha and Chatterjee 
studied electron acoustic periodic and solitary wave solutions in
unmagnetized \cite{r251} and magnetized \cite{r252, r253} quantum plasmas. They derived 
a KdV equation using the reductive perturbation method and investigated the 
associated nonlinear structures using 
bifurcation theory.

The manuscript is organized in the following way. In the next section, the
model and set of dynamic equations for studying nonlinear ion-acoustic waves
in unmagnetized quantum plasmas is presented. Using the reductive
perturbation method, the KdV equation is also derived with appropriate
boundary conditions. In Section III, the stationary wave solution is
obtained for ion-acoustic cnoidal waves using the Sagdeev potential
approach. In Section IV, the numerical analysis and plots are presented for
degenerate plasma cases at different plasma densities chosen from literature
and the conclusion is drawn in the final Section V.

%

\section{Basic Model and De\-ri\-va\-tion of Kor\-te\-weg-de Vries
E\-qua\-tion}

In order to study the electrostatic nonlinear periodic (cnoidal) waves in
unmagnetized electron-ion (EI) quantum plasmas, we will derive a
KdV equation using the reductive perturbation method.
A KdV equation for quantum ion-acoustic waves has already been derived by
Haas \textit{et al.} \cite{r3} with emphasis on localized solutions obtained under
decaying boundary conditions. For cnoidal waves, periodic boundary conditions 
are more appropriate, hence we will derive again the KdV equation
for quantum ion-acoustic waves. The set of dynamic equations for ion-acoustic wave
using QHD model is described as follows.

The ion continuity and momentum equations for ion fluid are given by

\begin{equation}
\frac{\partial n_{i}}{\partial t}+\frac{\partial }{\partial x}(n_{i}u_{i})=0,
\label{e1}
\end{equation}%
\begin{equation}
\frac{\partial u_{i}}{\partial t}+u_{i}\frac{\partial u_{i}}{\partial x}=-%
\frac{e}{m_{i}}\frac{\partial \phi }{\partial x}.  \label{e2}
\end{equation}%
The dynamic equation for the inertialess electron quantum fluid is described
by%
\begin{equation}
0=e\frac{\partial \phi }{\partial x}-\frac{1}{n_{e}}\frac{\partial p_{e}}{%
\partial x}+\frac{\hslash ^{2}}{2m_{e}}\frac{\partial }{\partial x}\left( 
\frac{1}{\sqrt{n_{e}}}\frac{\partial ^{2}}{\partial x^{2}}\sqrt{n_{e}}%
\right) .  \label{e3}
\end{equation}%
The Poisson equation is written as 
\begin{equation}
\frac{\partial ^{2}\phi }{\partial x^{2}}=\frac{e}{\varepsilon _{0}}%
(n_{e}-n_{i}),  \label{e4}
\end{equation}%
where $\phi $ is the electrostatic potential. The ion fluid density and
velocity are represented by $n_{i}$ and $u_{i}$ respectively, while $n_{e}$
is the electron fluid density. Also, $m_{e}$ and $m_{i}$ are the electron
and ion masses, $-e$ is the electronic charge, $\varepsilon _{0}$ and $%
\hslash $ are the dielectric and scaled Planck's constants. In equilibrium,
we have $n_{e0} = n_{i0} = n_{0}$ (say). Here $p_{e}$ is the electron pressure
and $p_{e}(n_{e})$ is obtained from the equation of state for the electron
fluid. The electrons are assumed to obey the equation of state pertaining to
one-dimensional zero-temperature Fermi gas \cite{r3,r4}, which is $%
p_{e} = m_{e}v_{Fe}^{2}n_{e}^{3}/3n_{0}^{2}$ where $n_{0}$ is the equilibrium
plasma density. Here $v_{Fe}$ is the Fermi velocity of electron, connected
to the Fermi temperature by $m_{e}v_{Fe}^{2}/2 = k_{B}T_{Fe}$ and $k_{B}$ is
the Boltzmann constant. The last term on the right hand side of the momentum
equations for electrons quantum fluid is the quantum force, which arises due
to the quantum Bohm potential and gives quantum diffraction or quantum
tunneling effects due to the wave-like nature of the charged particles. The
quantum effects due to ions are ignored in the model as they have large
inertia in comparison with the electrons.

In order to find the nonlinear ion-acoustic periodic waves in a quantum
plasma, the set of nonlinear dynamic equations are written in a normalized
form as follows,

\begin{equation}
\frac{\partial \tilde{n}_{i}}{\partial \tilde{t}}+\frac{\partial }{\partial 
\tilde{x}}(\tilde{n}_{i}\tilde{u}_{i})=0,  \label{e5}
\end{equation}%
\begin{equation}
\frac{\partial \tilde{u}_{i}}{\partial \tilde{t}}+\tilde{u}_{i}\frac{%
\partial }{\partial \tilde{x}}\tilde{u}_{i}=-\frac{\partial \Phi }{\partial 
\tilde{x}},  \label{e6}
\end{equation}%
\begin{equation}
0=\frac{\partial \Phi }{\partial \tilde{x}}-\frac{1}{2}\frac{\partial \tilde{%
n}_{e}^{2}}{\partial \tilde{x}}+\frac{H_{e}^{2}}{2}\frac{\partial }{\partial 
\tilde{x}}\left( \frac{1}{\sqrt{\tilde{n}_{e}}}\frac{\partial ^{2}}{\partial 
\tilde{x}^{2}}\sqrt{\tilde{n}_{e}}\right) ,  \label{e7}
\end{equation}%
\begin{equation}
\frac{\partial ^{2}\Phi }{\partial \tilde{x}^{2}}=\tilde{n}_{e}-\tilde{n}%
_{i}.  \label{e8}
\end{equation}%
The normalization of space, time, ion velocity and electrostatic potential
is defined by $\tilde{x}\rightarrow \omega _{_{pi}}x/c_{s}$, $\tilde{t}%
\rightarrow \omega _{_{pi}} t$, $\tilde{u}_{i}\rightarrow u_{i}/c_{s}$ and $%
\Phi =e\phi /2k_{B}T_{Fe}$ respectively, where the ion plasma frequency\ and
ion-acoustic speed are $\omega _{pi}=\left( n_{0}e^{2}/\varepsilon
_{0}m_{i}\right) ^{1/2}$ and$\ c_{s}=\left( 2k_{B}T_{Fe}/m_{i}\right) ^{1/2}$
respectively, and the non-dimensional quantum parameter for electrons is
defined as $H_{e}=\hbar \omega _{pe}/2k_{B}T_{Fe}$ i.e., the ratio of
electron plasmon energy to the Fermi energy, here $\omega _{pe}=\left(
n_{0}e^{2}/\varepsilon _{0}m_{e}\right) ^{1/2}$ is the electron plasma
frequency. The normalization of electron and ion fluid density is defined as 
$\tilde{n}_{j}=n_{j}/n_{0}$ ($j=e,i$). In the following, for simplicity we
will not use the tilde sign.

In order to find an nonlinear evolution equation, a stretching of
independent variables $x,$ $t$ is defined as follows \cite{r10},%
\[
\xi =\varepsilon^{1/2}(x-V_{0}t) \,, \quad \tau =\varepsilon^{3/2}t, 
\]%
where $\varepsilon $ is a small parameter and $V_{0}$ is the phase velocity
of the wave to be determined later on. The perturbed quantities can be
expanded in the powers of $\varepsilon $,

\[
n_{i1}=1+\varepsilon n_{i1}+\varepsilon ^{2}n_{i2}+..., 
\]%
\[
n_{e1}=1+\varepsilon n_{e1}+\varepsilon ^{2}n_{e2}+..., 
\]%
\[
u_{i}=\varepsilon u_{i1}+\varepsilon ^{2}u_{i2}+..., 
\]%
\begin{equation}
\Phi =\varepsilon \Phi _{1}+\varepsilon ^{2}\Phi _{2}+...  \label{e9}
\end{equation}%
Moreover, $\partial /\partial x=\varepsilon ^{1/2}\, \partial /\partial \xi $
and $\partial /\partial t=\varepsilon ^{3/2}\partial /\partial \tau
-V_{0}\varepsilon ^{1/2}\partial /\partial \xi $.

From ion continuity and momentum equations the lowest order $(\sim
\varepsilon ^{3/2})$ terms gives 
\begin{equation}
-V_{0}\frac{\partial n_{i1}}{\partial \xi }+\frac{\partial u_{i1}}{\partial
\xi }=0,  \label{e10}
\end{equation}%
\begin{equation}
-V_{0}\frac{\partial u_{i1}}{\partial \xi }+\frac{\partial \Phi _{1}}{%
\partial \xi }=0,  \label{e11}
\end{equation}%
\begin{equation}
\frac{\partial \Phi _{1}}{\partial \xi }-\frac{\partial n_{e1}}{\partial \xi 
}=0.  \label{e12}
\end{equation}%
The lowest order $(\sim \varepsilon )$ term of Poisson equation gives%
\begin{equation}
n_{i1}=n_{e1}.  \label{e13}
\end{equation}%
Multiplying Eq. (\ref{e10}) by $V_{0}$ and adding with Eq. (\ref{e11}), we
have 
\begin{equation}
\frac{\partial n_{i1}}{\partial \xi }=\frac{1}{V_{0}^{2}}\frac{\partial \Phi
_{1}}{\partial \xi }.  \label{e14}
\end{equation}%
Using Eqs. (\ref{e12}), (\ref{e13}) and (\ref{e14}), we have 
\begin{equation}
V_{0}=\pm 1.  \label{e16}
\end{equation}%
which is the normalized phase velocity of the ion-acoustic wave. From now
on, we set $V_{0}=1$ without loss of generality.

Now integrating Eqs. (\ref{e11}), (\ref{e12}), (\ref{e14}) and using (\ref%
{e13}) we have%
\begin{equation}
n_{i1}=n_{e1}=\Phi _{1}+c_{1}(\tau )  \label{e17}
\end{equation}%
and 
\begin{equation}
u_{i1}=\Phi _{1}+c_{2}(\tau ),  \label{e18}
\end{equation}%
where $c_{1}(\tau )$ and $c_{2}(\tau )$ are at this point arbitrary
functions of $\tau $ only.

Now collecting the next higher order terms from ion dynamic equations, we
have%
\begin{equation}
-\frac{\partial n_{i2}}{\partial \xi }+\frac{\partial u_{i2}}{\partial \xi }+%
\frac{\partial n_{i1}}{\partial \tau }+\frac{\partial }{\partial \xi }%
(n_{i1}u_{i1})=0,  \label{e19}
\end{equation}%
\begin{equation}
-\frac{\partial u_{i2}}{\partial \xi }+\frac{\partial \Phi _{2}}{\partial
\xi }+\frac{\partial u_{i1}}{\partial \tau }+u_{i1}\frac{\partial u_{i1}}{%
\partial \xi }=0,  \label{e20}
\end{equation}%
\begin{equation}
\frac{\partial \Phi _{2}}{\partial \xi }-\frac{\partial n_{e2}}{\partial \xi 
}-\frac{1}{2}\frac{\partial }{\partial \xi }n_{e1}^{2}+\frac{H_{e}^{2}}{4}%
\frac{\partial ^{3}n_{e1}}{\partial \xi ^{3}}=0.  \label{e21}
\end{equation}%
The next higher order ($\sim \varepsilon ^{2}$) term of Poisson equation
gives%
\begin{equation}
\frac{\partial ^{2}\Phi _{1}}{\partial \xi ^{2}}=n_{e2}-n_{i2}.  \label{e22}
\end{equation}%
Adding Eqs. (\ref{e19}) and (\ref{e20}), we obtain%
\begin{equation}
\frac{\partial n_{i2}}{\partial \xi }=\frac{\partial n_{i1}}{\partial \tau }+%
\frac{\partial }{\partial \xi }(n_{i1}u_{i1})+\frac{\partial u_{i1}}{%
\partial \tau }+u_{i1}\frac{\partial }{\partial \xi }u_{i1}+\frac{\partial
\Phi _{2}}{\partial \xi },  \label{e23}
\end{equation}%
and from Eq. (\ref{e21}) we have 
\begin{equation}
\frac{\partial n_{e2}}{\partial \xi }=\frac{\partial \Phi _{2}}{\partial \xi 
}-\frac{1}{2}\frac{\partial }{\partial \xi }n_{e1}^{2}+\frac{H_{e}^{2}}{4}%
\frac{\partial ^{3}n_{e1}}{\partial \xi ^{3}}.  \label{e24}
\end{equation}%
Using $n_{i1}$, $u_{i1}$ and $n_{e1}$ from Eqs. (\ref{e17}) and (\ref{e18})
in Eqs. (\ref{e23}) and (\ref{e24}), we obtain

\begin{equation}
\frac{\partial n_{i2}}{\partial \xi }=2\frac{\partial \Phi _{1}}{\partial
\tau }+3\Phi _{1}\frac{\partial \Phi _{1}}{\partial \xi }+\left(
2c_{2}+c_{1}\right) \frac{\partial \Phi _{1}}{\partial \xi }+\frac{\partial
\Phi _{2}}{\partial \xi }  \label{e25}
\end{equation}%
and 
\begin{equation}
\frac{\partial n_{e2}}{\partial \xi }=-\Phi _{1}\frac{\partial \Phi _{1}}{%
\partial \xi }-c_{2}\frac{\partial \Phi _{1}}{\partial \xi }+\frac{H_{e}^{2}%
}{4}\frac{\partial ^{3}\Phi _{1}}{\partial \xi ^{3}}+\frac{\partial \Phi _{2}%
}{\partial \xi }.  \label{e26}
\end{equation}%
Applying periodic boundary conditions we get $\partial c_{1}/\partial \tau
=\partial c_{2}/\partial \tau =0$, so that the functions $c_{1} $ and $c_{2}$
become independent of both $\xi $ and $\tau $ and are from now on constants.

Differentiating Eq. (\ref{e22}) and using Eqs. (\ref{e25}) and (\ref{e26}),
and after some simplifications, we have the KdV equation for the nonlinear
dynamics of ion-acoustic waves in a quantum plasma as follows,%
\begin{equation}
\frac{\partial \Phi }{\partial \tau }+2\Phi \frac{\partial \Phi }{\partial
\xi }+c_{0}\frac{\partial \Phi }{\partial \xi }+D\frac{\partial ^{3}\Phi }{%
\partial \xi ^{3}}=0,  \label{e27}
\end{equation}%
where%
\begin{eqnarray}
c_{0} &=&c_{1}+c_{2},  \nonumber \\
D &=&\frac{1}{2}\left( 1-\frac{H_{e}^{2}}{4}\right) .  \label{e28}
\end{eqnarray}%
Here $\Phi _{1}$ has been replaced by $\Phi $.

In the above KdV equation (\ref{e27}), the term containing an arbitrary
constant $c_{0}$ can be removed with a Galilean transformation. Hence we can
set $c_{0}=0$ without loss of generality. It can be noticed easily from
dispersive coefficient $D$ that the cnoidal wave solution exist only when $%
H_{e}\neq 2$, so that the dispersive coefficient does not disappear to
balance the nonlinearity.


\section{Nonlinear Periodic Wave Solutions}

In order to find the steady state cnoidal and solitary waves solutions of
the KdV Eq. (\ref{e27}) for quantum ion-acoustic waves, we follow the same
procedure as already done in Refs.\cite{r24,r25}. Assume a solution $\Phi
(\eta )$, where $\eta =\xi -u\tau $ and $u$ is the velocity of the nonlinear
structure moving with the frame. Therefore, Eq. (\ref{e27}) can be written as

\begin{equation}
D\frac{d^{3}\Phi }{d\eta ^{3}}+\frac{d}{d\eta }\left( \Phi ^{2}-u\Phi
\right) =0.  \label{e29}
\end{equation}%
As said, the arbitrary constant $c_{0}$ has been ignored, since it gives
just a shift in the velocity of the nonlinear structure.

After integration of Eq. (\ref{e29}), we get the equation of a conservative
nonlinear oscillator i.e., 
\begin{equation}
\frac{d^{2}\Phi }{d\eta ^{2}}=-\frac{dW}{d\Phi },  \label{e30}
\end{equation}%
where its potential energy $W = W(\Phi)$ is defined as%
\begin{equation}
W(\Phi )=\frac{1}{3D}\Phi ^{3}-\frac{u}{2D}\Phi ^{2}+\rho _{0}\Phi .
\label{e31}
\end{equation}%
Here $\rho _{0}$ is an integration constant. The potential function $W(\Phi
) $ has two points of extremum for $\Phi $ i.e. $\Phi =\tilde{\Phi}_{1,2}$
defined by $\partial W/\partial \Phi =0$, which are given by

\begin{equation}
\tilde{\Phi}_{1,2}=\frac{u}{2}\pm \sqrt{\frac{u^{2}}{4}-D\rho _{0}}.
\label{e32}
\end{equation}%
Thus, there are two equilibrium states. One of them defines a saddle point
while the other one represents a center point i.e., the bottom of the
potential well \cite{r24}. Moreover $u^{2}/4-D\rho _{0}>0$ must hold for
real values. The zero's of the potential energy (\ref{e31}) i.e., $\Phi
=z_{1},z_{2},z_{3}$ are given as follows, 
\begin{equation}
z_{1}=0 \,, \quad z_{2,3}=\frac{3}{2}\left( \frac{u}{2}\pm \sqrt{\frac{u^{2}%
}{4}-\frac{4}{3}D\rho _{0}}\right) .  \label{e34}
\end{equation}%
To get the shape of the real potential well $u^{2}/4-4D\rho _{0}/3>0$ must
holds. Note that the potential well having the center for $\Phi >0$
(positive) defines the compressive cnoidal waves and solitons, while in the
case of $\Phi <0$ (negative) the potential well defines the rarefactive
cnoidal waves and solitons \cite{r24}. The shape of the potential well
strongly depends on the sign of dispersive coefficient $D$ (see Eq. (\ref{e31}%
)). As described in Eq. (\ref{e28}), the value of the dispersive coefficient $%
D$ in the KdV equation is positive for electron quantum parameter $H_{e}<2$
case, while it becomes negative for $H_{e}>2$. Therefore, the formation of
compressive or rarefactive ion-acoustic nonlinear structure depends on the
value of the quantum parameter $H_{e}$. The amplitude of the nonlinear
structures is defined by the width of the potential well, which is the
length between the last zero of the potential well and the saddle point (see
Eqs. (\ref{e32}) and (\ref{e34})).

The energy first integral associated to (\ref{e30}) is 
\begin{equation}
\frac{1}{2}\left( \frac{d\Phi }{d\eta }\right) ^{2}+W(\Phi )=\frac{E_{0}^{2}%
}{2},  \label{e35}
\end{equation}%
where $E_{0}^{2}$ is the integration constant (assumed positive definite in
order to access cnoidal wave solutions). 

Using Eq. (\ref{e31}) in (\ref{e35}), we have%
\begin{equation}
\left( \frac{d\Phi }{d\eta }\right) ^{2}=E_{0}^{2}-\frac{2}{3D}\Phi ^{3}+%
\frac{u}{D}\Phi ^{2}-2\rho _{0}\Phi .  \label{e36}
\end{equation}%
Let us consider the initial conditions $\Phi (0)=\Phi _{0}$ and $d\Phi
(0)/d\eta =0$. Then we can define%
\begin{equation}
E_{0}^{2}=\frac{2}{3D}\Phi _{0}^{3}-\frac{u}{D}\Phi _{0}^{2}+2\rho _{0}\Phi
_{0}.  \label{e37}
\end{equation}%
Using Eq. (\ref{e37}) in Eq.  (\ref{e36}) and after factorization, we have 
\begin{equation}
\left( \frac{d\Phi }{d\eta }\right) ^{2}=\frac{2}{3D}(\Phi _{0}-\Phi )(\Phi
-\Phi _{1})(\Phi -\Phi _{2}),  \label{e38}
\end{equation}%
where 
\begin{equation}
\Phi _{1,2}=\frac{3}{2}\left[ \frac{u}{2}-\frac{\Phi _{0}}{3}\pm \sqrt{\frac{%
1}{3}(b_{1}-\Phi _{0})(\Phi _{0}-b_{2})}\right] ,  \label{e39}
\end{equation}%
and%
\begin{equation}
b_{1,2}=\frac{u}{2}\pm 2\sqrt{\frac{u^{2}}{4}-D\rho _{0}}.  \label{e40}
\end{equation}%
In addition, the following inequalities should be kept: $b_{2}\leq \Phi
_{0}\leq b_{1}$ or $b_{1}\leq \Phi _{0}\leq b_{2}$. From Eqs. (\ref{e36})-(%
\ref{e38}), we have the following relation,%
\begin{equation}
u=\frac{2}{3}(\Phi _{0}+\Phi _{1}+\Phi _{2}).  \label{e41}
\end{equation}%
The periodic (cnoidal) wave solution of Eq. (\ref{e38}) is given \cite{r26} by 
\begin{equation}
\Phi (\eta )=\Phi _{1}+(\Phi _{0}-\Phi _{1})\mathrm{cn}^{2}(R\eta ,s),
\label{e42}
\end{equation}%
where cn is the Jacobian elliptic function, $s$ is the modulus defined as 
\begin{equation}
s^{2}=\frac{(\Phi _{0}-\Phi _{1})}{(\Phi _{0}-\Phi _{2})}<1,  \label{e43}
\end{equation}%
and the quantity $R=\sqrt{\frac{1}{6D}(\Phi _{0}-\Phi _{2})}$.

The amplitude $A$ of the cnoidal wave is defined from Eq. (\ref{e42}) as
follows,%
\begin{equation}
A=\Phi _{0}-\Phi _{1}.  \label{e44}
\end{equation}%
As it is seen from the solution (\ref{e42}) at $\eta =0$, we have the
initial condition $\Phi (0)=\Phi _{0}$. In addition, the real numbers $\Phi
_{i}$ $(i=0,1,2)$ are ordered as $\Phi _{0}>\Phi _{1}\geq \Phi _{2}$ and $%
\Phi _{1}\leq \Phi \leq \Phi _{0}$ for $D>0$.

The modulus $0<s<1$\ is a measure of the nonlinearity of the wave. The case $%
s<<1$ corresponds to the weakly nonlinear oscillations near the bottom of
the potential well and\ the elliptic functions are close to trigonometric
ones. At $s\rightarrow 0$, the expression (\ref{e42})\ passes to solution of
linear equations \cite{r26}.

The wavelength $\lambda $ of the cnoidal waves is defined as 
\begin{equation}
\lambda =4\sqrt{\frac{3D}{2(\Phi _{0}-\Phi _{2})}}K(s),  \label{e45}
\end{equation}%
where $K(s)$ is the complete elliptic integral of the first kind and the
corresponding frequency is $f=v_{1}/\lambda $ (where $v_{1}$ is the velocity
of the cnoidal waves). The velocity $v_{1}$ of the cnoidal waves in the
laboratory frame is equal to $v_{1}=V_{0}+u$, where the expression for the
frame velocity $u$ is given by,%
\begin{equation}
u=\frac{2}{3}\frac{(\Phi _{0}-\Phi _{1})}{s^{2}}(2-s^{2})+2\Phi _{2},
\label{e46}
\end{equation}%
which has been obtained using the expression of the modulus $s$ described by
Eq. (\ref{e43}).

The mean value of $\Phi $ can be expressed as%
\begin{equation}
\bar{\Phi}=\frac{1}{\lambda }\int_{0}^{\lambda}\Phi (\eta )d\eta =\Phi
_{2}+(\Phi _{0}-\Phi _{2})\frac{E(s)}{K(s)},  \label{e47}
\end{equation}%
where $E(s)$ is the complete integral of the second kind.

The limiting case of the soliton i.e. $s=1$, can be obtained at $\Phi
_{0}\approx b_{1}$ or $\Phi _{0}\approx b_{2}$ (see Eq. (\ref{e40})), so that%
\begin{equation}
\Phi _{1}\approx \Phi _{2}=\frac{u}{2}\mp \sqrt{\frac{u^{2}}{4}-D\rho _{0}}.
\label{e48}
\end{equation}%
Further, we take into account, 
\begin{equation}
K(s)\approx \frac{1}{2}\mathrm{ln}\left( \frac{16}{1-s^{2}}\right)
\rightarrow \infty \,,\quad \mathrm{cn\,z}\rightarrow \frac{1}{\mathrm{ch\,z}%
},  \label{e49}
\end{equation}%
which imply that the wavelength of the cnoidal waves defined in (\ref{e45})
tends to infinity and the solution (\ref{e42}) passes to a soliton-like
shape \cite{r26} i.e.,%
\begin{equation}
\Phi (\eta )=\Phi _{1}+\frac{A}{\mathrm{ch}^{2}\left( \sqrt{\frac{1}{6D}%
(\Phi _{0}-\Phi _{1})}\eta \right) },  \label{e50}
\end{equation}%
where $\triangle =\sqrt{6\left \vert D/A\right \vert }$ is the width of the
soliton and $A$ is its amplitude, defined by Eq. (\ref{e44}). As it follows
from Eq. (\ref{e46}), the propagation velocity of the solitons becomes%
\begin{equation}
u=2\Phi _{1}+\frac{2}{3}A.  \label{e51}
\end{equation}%
From Eq. (\ref{e50}), we see that $\Phi _{1}$ defines the potential at $\eta
\rightarrow \pm \infty $.

Thus, at large values of $s$ i.e., $s\rightarrow 1$, $A=const$ the periodic
wave asymptotically approaches to the sequence of solitons having the
amplitude $A$ (relatively to the level $\Phi =\bar{\Phi}=\Phi _{1}$). By the
order of magnitude, the distance between them is equal to 
\begin{equation}
\lambda =\bigtriangleup \left \vert \mathrm{ln}(1-s^{2})\right \vert ,
\label{e52}
\end{equation}%
where $\bigtriangleup $ is the width of the soliton already defined above.


\section{Numerical Analysis}

The numerical plots of the nonlinear wave potential and the phase plane
plots of the cnoidal wave structures and solitons are shown in Figs.1-3 at
different densities for a degenerate electron plasma cases, such as
astrophysical plasmas, laser plasmas and ultra-cold plasmas. For a
completely degenerate electron plasma, the electron Fermi energy and density
are related as $k_{B}T_{Fe}=\hslash ^{2}(3\pi ^{2}n_{0})^{2/3}/2m_{e}$ and
the Fermi temperature of degenerate electrons ($T_{Fe}$) should be much
greater than the thermal temperature $T$ of the system i.e., $T \ll T_{Fe}$.
The quantum parameter for electrons is related to density as $H_{e}\sim
n_{0}^{-1/6}$, which shows that in a completely degenerate electron plasma
case, the value of the quantum parameter decreases with the increase in the
plasma density. So the quantum diffraction effects tend to be less relevant
in dense plasmas. In case of astrophysical plasma conditions i.e., $%
n_{0}=10^{36}m^{-3}$, the quantum parameter for degenerate electrons comes
out to be $H_{e}=0.05$ and the condition for thermal temperature becomes $%
T<<10^{9}K$, while for laser plasmas we have $n_{0}=10^{32}m^{-3}$ then $%
H_{e}=0.24$ and $T<<10^{7}K$. Further, for ultra-cold plasmas, we have $%
n_{0}=2.7\times 10^{26}m^{-3}$ for which $H_{e}=2.002$ and $T<<1800K$ \cite%
{r28,r29}.

\begin{figure}[h]
 \includegraphics[width=5.0in,height=5.0in]{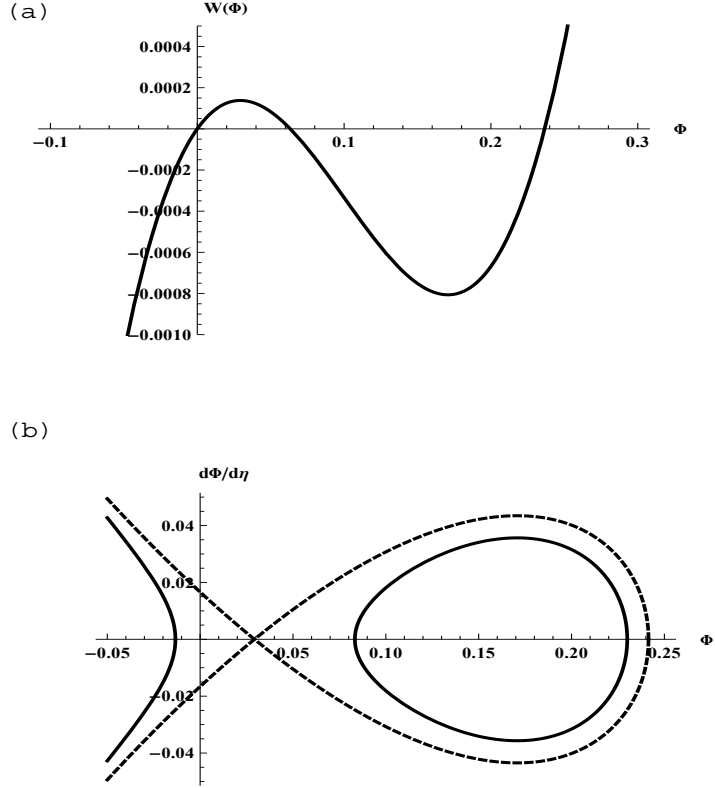}
  \caption{(a) The nonlinear potential $W(\Phi )$ is plotted for $%
\rho _{0}=0.01$, $u=0.2$ and $H_{e}=0.05$. (b) The phase plane plots of the
compressive ion-acoustic cnoidal wave (solid bounded curve) and soliton
(dotted curve) are shown for the same numerical values as in Fig. 1a.
Dimensionless variables are used.}
 \label{f1}
\end{figure}

\begin{figure}[h]
 \includegraphics[width=5.0in,height=5.0in]{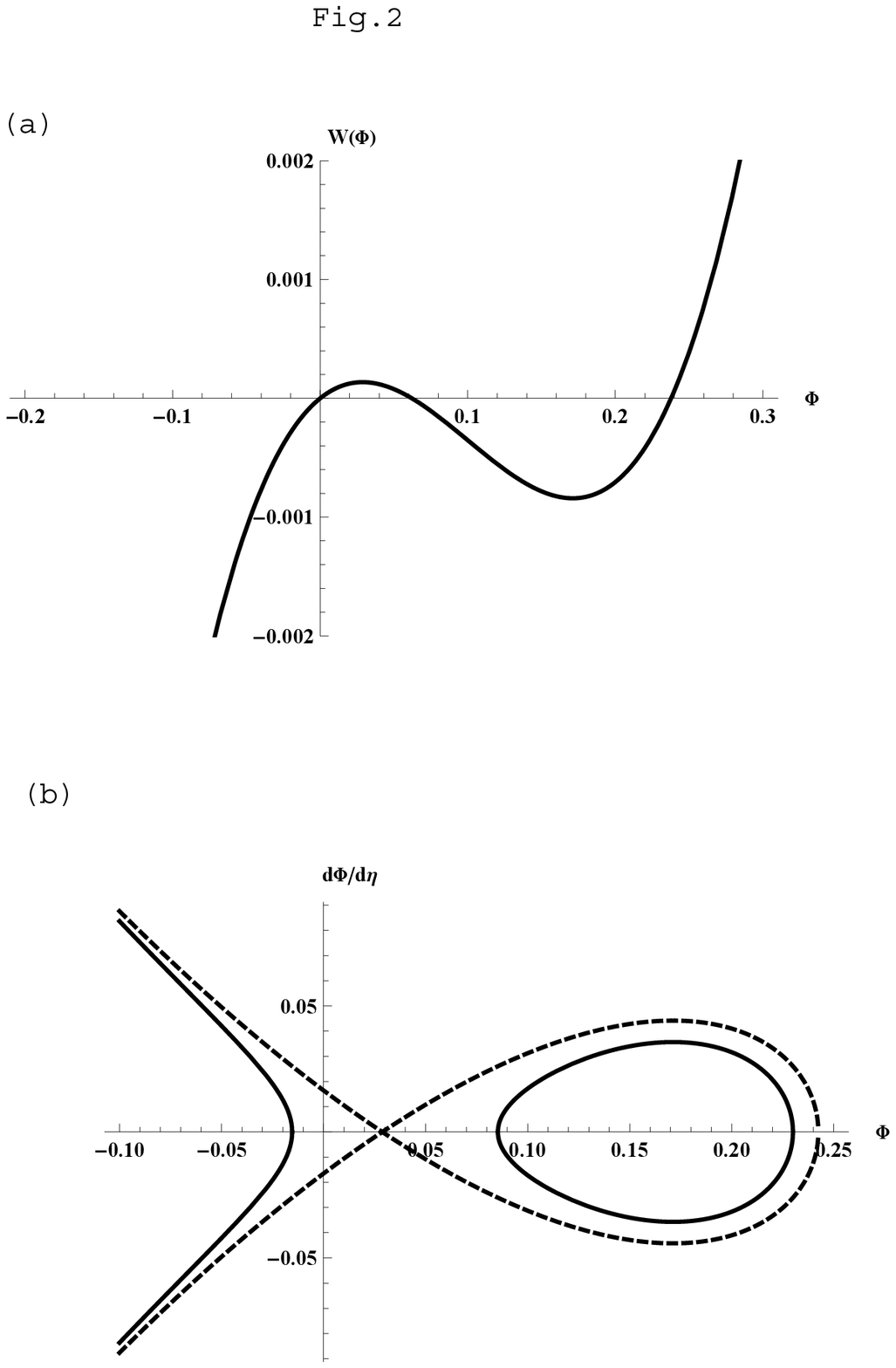}
  \caption{(a) The nonlinear potential $W(\Phi )$ is plotted for $%
\rho _{0}=0.01$, $u=0.2$ and $H_{e}=0.24$ (b) The phase plane plots of the
compressive ion-acoustic cnoidal wave (solid bounded curve) and soliton
(dotted curve) are shown for the same numerical values as in Fig. 2a.
Dimensionless variables are used.}
 \label{f2}
\end{figure}

\begin{figure}[h]
 \includegraphics[width=5.0in,height=5.0in]{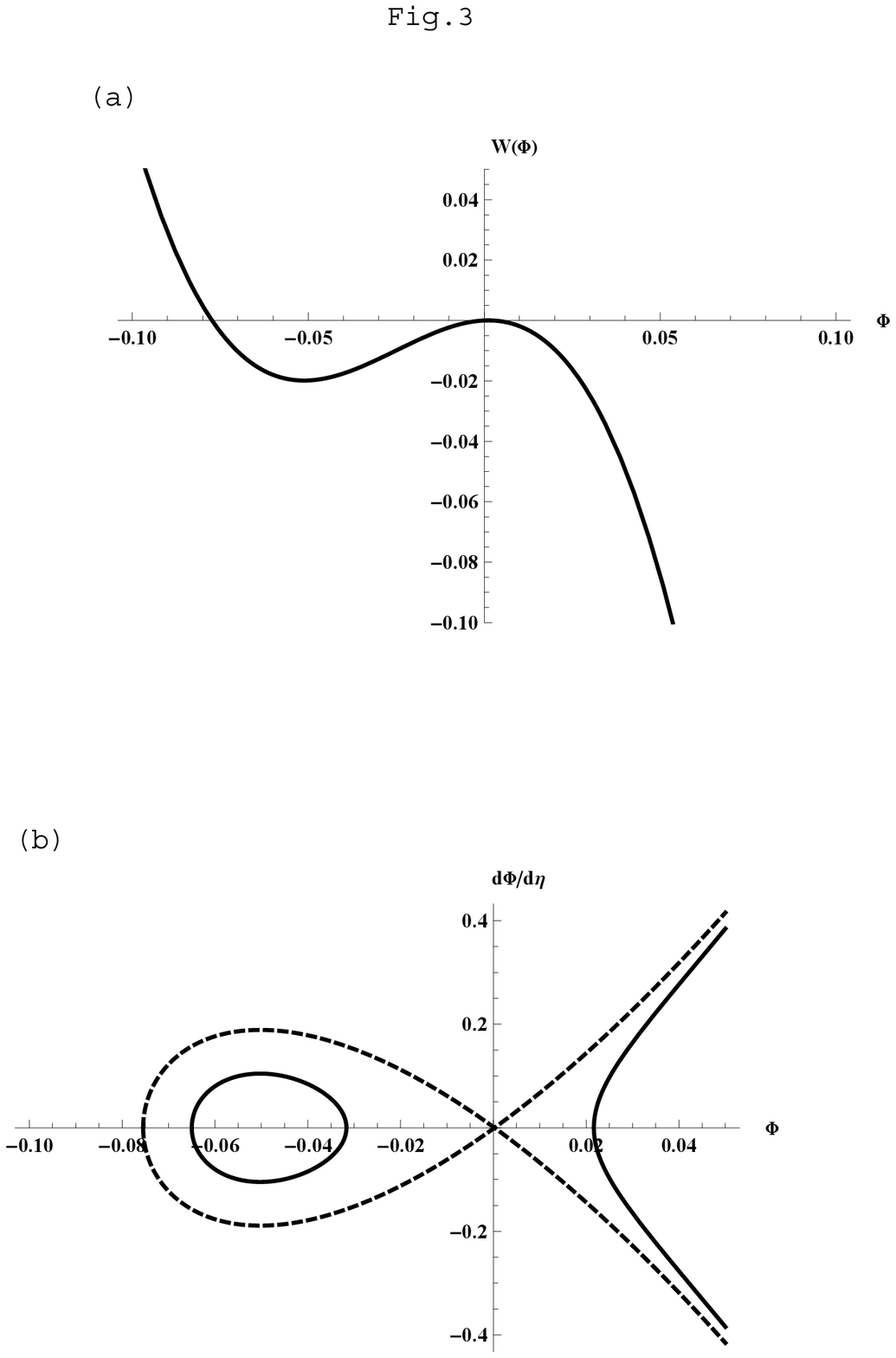}
  \caption{(a) The nonlinear potential $W(\Phi )$ is plotted for $%
\rho _{0}=0.01$, $u=-0.05$ and $H_{e}=2.002$ (b) The phase plane plots of
the rarefactive ion-acoustic cnoidal wave (solid bounded curve) and soliton
(dotted curve) are shown for the same numerical values as in Fig. 3a.
Dimensionless variables are used.}
 \label{f3}
\end{figure}

The formation of ion-acoustic compressive (rarefactive) nonlinear structures
depends on the value of the quantum parameter for electrons i.e., $H_{e}<2$ (%
$H_{e}>2$). It is also noticed that the velocity of the nonlinear structure
is positive i.e., $u>0$ for compressive cnoidal waves and solitons with
quantum parameter for degenerate electrons lies in the range $0\leq H_{e}<2$%
. For the $H_{e}>2$ case, rarefactive ion-acoustic cnoidal waves and
solitons structures are formed and its solution exist only when the velocity
of the nonlinear structure is negative $u<0$ i.e., it moves in the backward
direction.

It can be seen from Figs. 1a and 2a that the Sagdeev potential $W(\Phi)$
are formed for $\Phi > 0$ for the electron quantum parameter values $%
H_{e}=0.05$ and $0.24$. The corresponding compressive ion-acoustic cnoidal
wave (solid curve) and solitons (dotted curve) structures are shown in the
phase plane plots of Fig. 1b and 2b respectively. The cnoidal wave structure
(solid bounded curve) is formed inside the separatrix (dotted curve) which
represent a soliton structure as shown in the figures. The Sagdeev potential
plot $W(\Phi )$ is formed with $\Phi <0$ for the electron quantum parameter
value $H_{e}=2.002$ ($>2$) as shown in the Fig. 3a. The corresponding
rarefactive ion-acoustic cnoidal wave (solid bounded curve) and soliton
(dotted curve) structures in the phase plane plot are shown in the Fig. 3b.
Rarefactive nonlinear ion-acoustic cnoidal wave (solid bounded curve)
structures are also formed inside the separatrix (dotted curve), which
represent the soliton.

The plots of the compressive ion-acoustic cnoidal wave from Eq. (\ref%
{e42}) and solitons from Eq. (\ref{e50}) for $H_{e} < 2$ case are
shown in the Figs. 4a and 4b respectively. It can be seen from the Fig. 4a
that there is a little decrease in the wavelength (frequency) of the
compressive ion-acoustic cnoidal wave case with the increase in the quantum
parameter of the degenerate electrons. The little increase in the width of
the compressive ion-acoustic solitons with the increase in the value of
quantum parameter is shown in the Fig. 4b. Similarly, the plots for the
rarefactive ion-acoustic cnoidal wave and soliton structures for $H_{e}>2$%
case are shown in the Figs. 5a and 5b respectively. It can be seen
from the Fig. 5a that wavelength (frequency) of the rarefactive ion-acoustic
cnoidal wave increases and its amplitude decreases significantly with the
little increase in the value of quantum parameter for $H_{e} > 2$ 
case. Also, the decrease in the wave amplitude as well as increase in the
width of the rarefactive ion-acoustic soliton with the increase in the value
of quantum parameter is shown in Fig. 5b.

\begin{figure}[h]
 \includegraphics[width=5.0in,height=5.0in]{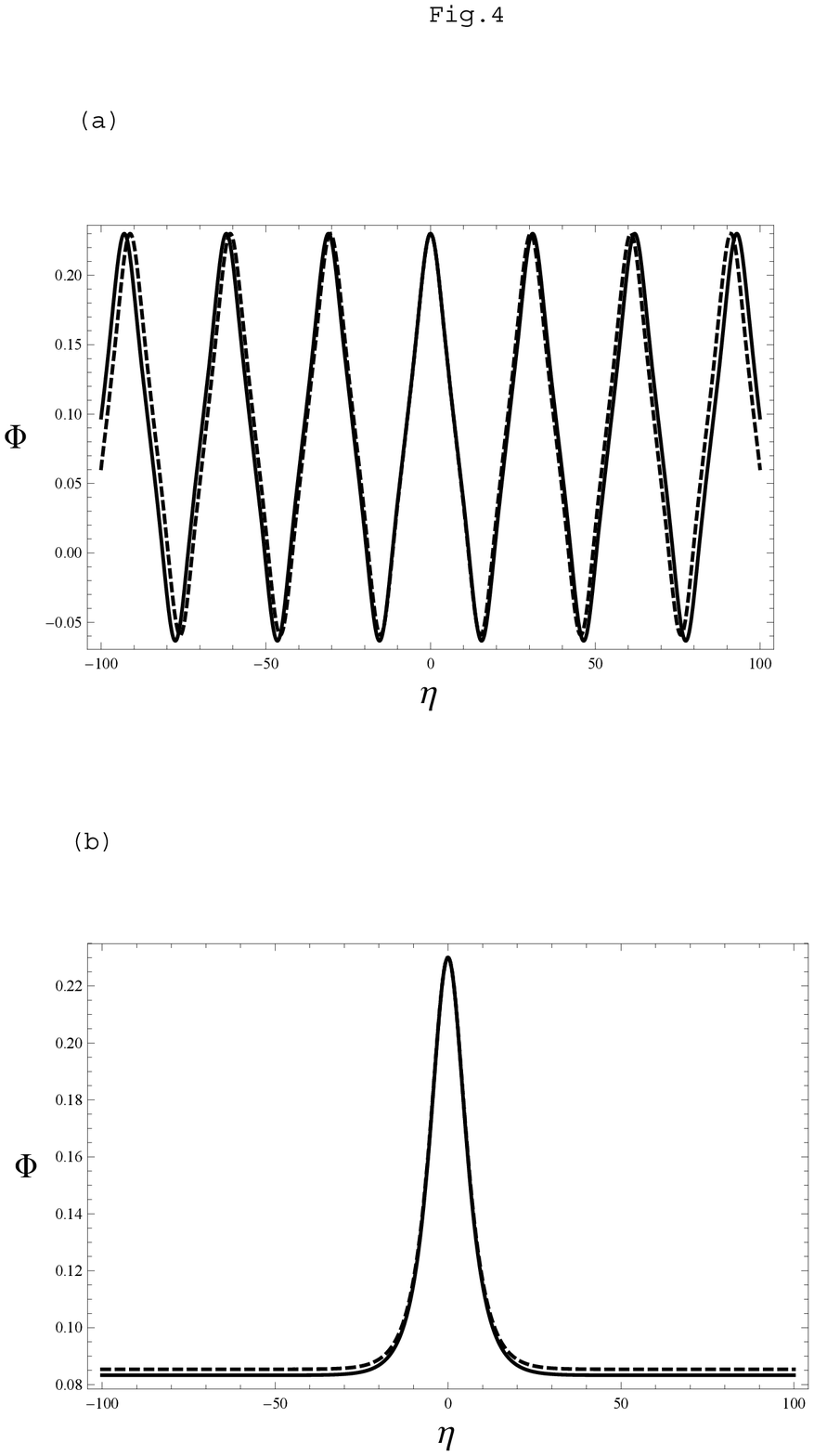} 
  \caption{(a) The plots of compressive ion-acoustic cnoidal waves
(periodic wave oscillations) from Eq.  (\ref{e42}) and (b) solitons (single
pulse) from Eq. (\ref{e50}) are shown for the $H_{e} < 2$ case i.e., with 
$H_{e} = 0.05$ (solid) and $H_{e} = 0.24$ (dotted) respectively. Dimensionless variables are used.}
 \label{f4}
\end{figure}

\begin{figure}[h]
 \includegraphics[width=5.0in,height=5.0in]{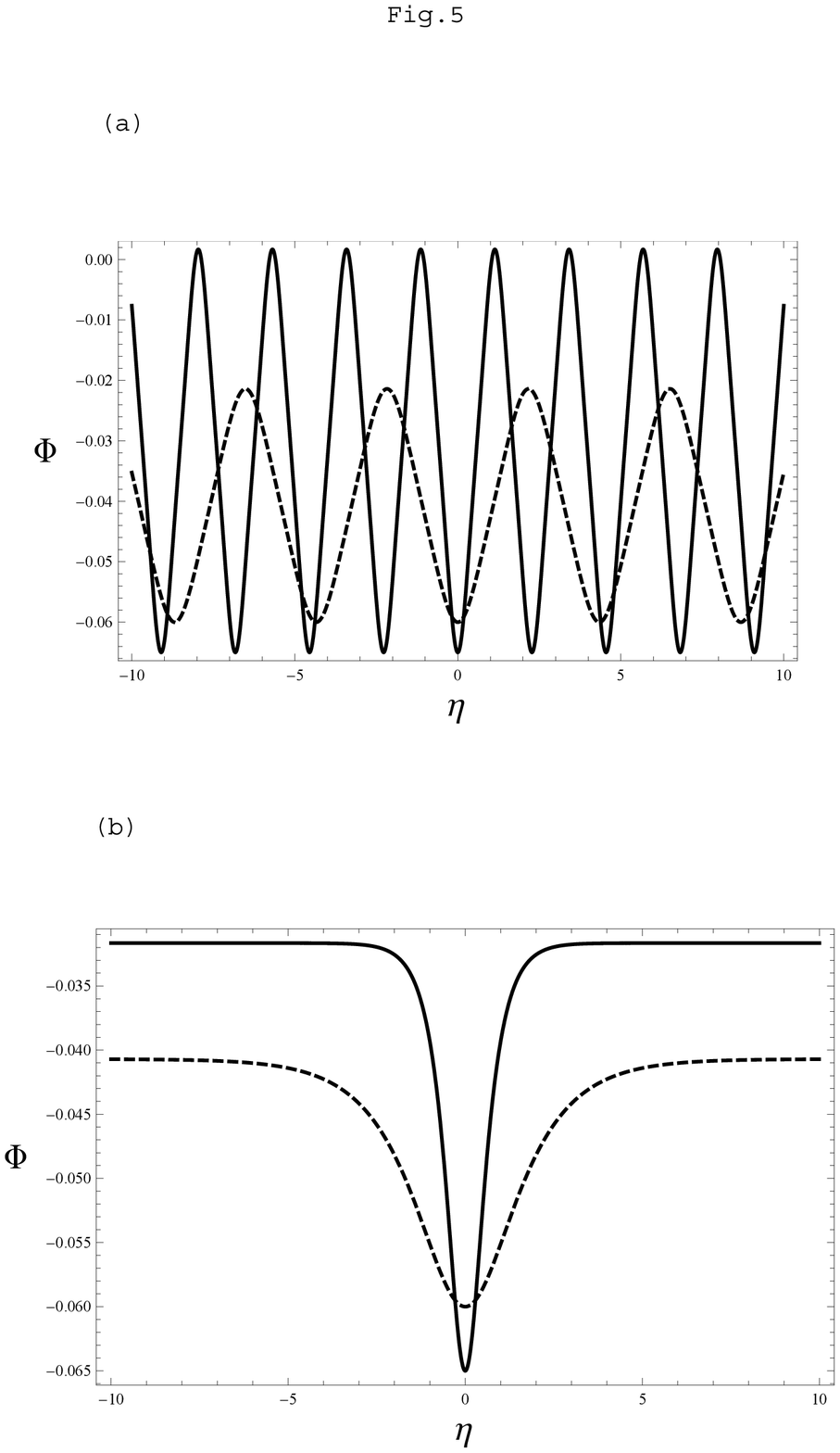}
  \caption{(a) The plots of compressive ion-acoustic cnoidal waves
(periodic wave oscillations) from Eq.  (\ref{e42}) and (b) solitons (single
pulse) from Eq. (\ref{e50}) are shown for the $H_{e} > 2$ case i.e., with 
$H_{e} = 2.002$ (solid) and $H_{e} = 2.01$ (dotted) respectively. Dimensionless variables are used.}
 \label{f5}
\end{figure}

The variations of the wavelength and frequency with wave
amplitude of the compressive ion-acoustic cnoidal waves at different quantum
parameters are shown in Fig. 6a and 6b, respectively. It
can be seen from the figures that the wavelength increases, while the
frequency decreases with the increase in the amplitude. Also, the decrease
in the wavelength and increase in the wave frequency is found with the
increase in the electron quantum parameter for compressive ($H_{e}<2$)
ion-acoustic cnoidal waves case. However, this decrease in the wavelength
and increase in the wave frequency with the wave amplitude for the
ion-acoustic compressive cnoidal waves case seems to be very small at the
chosen degenerate plasma densities as indicated in Figs. 6a and 
6b. The dependence of the wavelength and frequency on the wave
amplitude in case of rarefactive ion-acoustic cnoidal waves case i.e., at
quantum parameters with values $H_{e}>2$ are shown in Figs. 7a and 
7b, respectively. The wavelength increases while the frequency
decreases with the increase in the amplitude of the rarefactive ion-acoustic
cnoidal waves in quantum plasmas. On the other hand, the wavelength is found
to be increased, while frequency decreases for rarefactive ion-acoustic
cnoidal waves case with the increase in the value of the quantum parameter
for degenerate electrons as shown in the Figs. 7a and 7b.

\begin{figure}[h]
 \includegraphics[width=5.0in,height=5.0in]{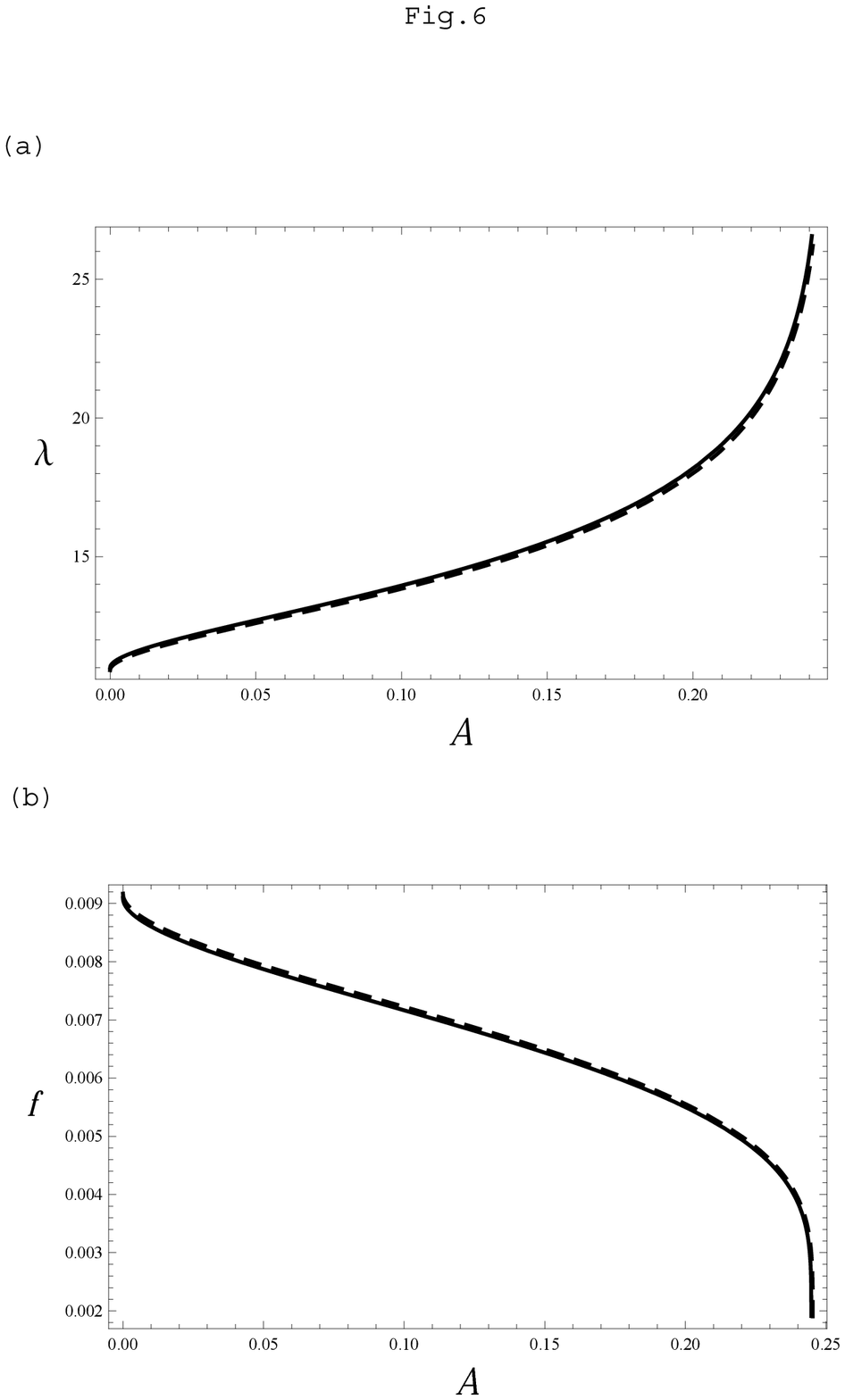}
  \caption{(a) The dependence of wavelength $\lambda$ on compressive
ion-acoustic cnoidal wave amplitude $A$ is shown for degenerate electron
quantum parameter $H_{e}=0.05$ (solid curve) and $H_{e}=0.24$ (dotted curve)
with $v_{1}=0.1$.  (b) The dependence of frequency $f$ on compressive
ion-acoustic cnoidal wave amplitude $A$ with same parameters as in Fig. 4a. Dimensionless variables are used.}
 \label{f6}
\end{figure}


\begin{figure}[h]
 \includegraphics[width=5.0in,height=5.0in]{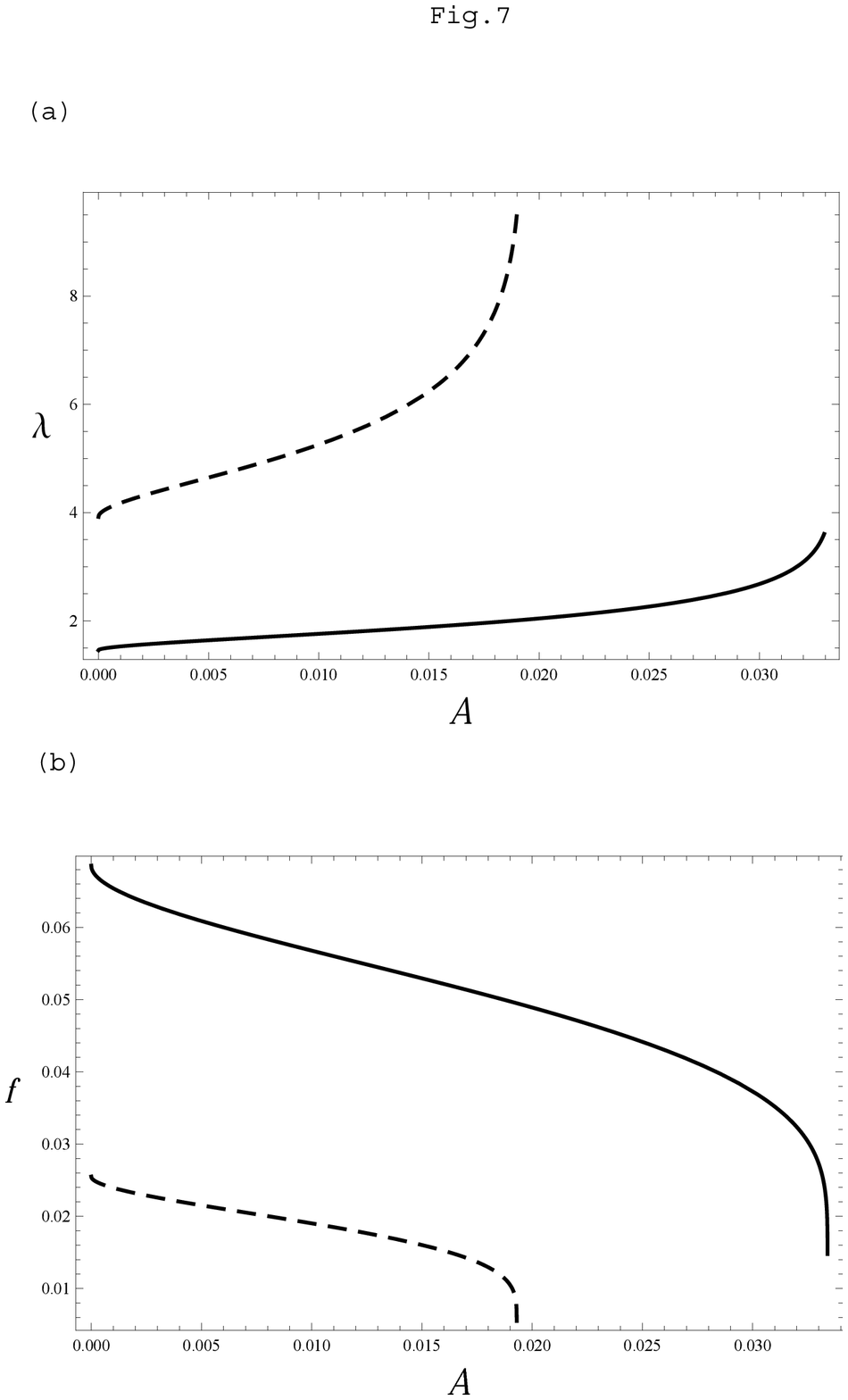}
  \caption{(a) The dependence of wavelength $\lambda $ on
rarefactive ion-acoustic cnoidal wave amplitude $A$ is shown for degenerate
electron quantum parameter $H_{e} = 2.002$ (solid curve) and $H_{e} = 2.01$
(dotted curve). (b) The dependence of frequency $f$ on rarefactive
ion-acoustic cnoidal wave amplitude $A$ with the same parameters as in Fig. 6a. Dimensionless variables are used.}
 \label{f7}
\end{figure}


\section{Conclusion}

To conclude, we have studied for the first time the ion-acoustic cnoidal
waves and solitons in an unmagnetized quantum plasma. The KdV equation for
ion-acoustic waves in a quantum plasma was obtained using the reductive
perturbation method with periodic wave boundary conditions, appropriate to
study cnoidal waves. It is found that both compressive and rarefactive
nonlinear ion-acoustic cnoidal wave structures are formed in such a
degenerate plasma, which depends on the quantum parameter i.e., $%
H_{e}\gtrless 2$. The dependence of wave frequency and wavelengths on the
nonlinear ion-acoustic wave amplitude is also investigated at different
values of quantum parameters with the degenerate plasma densities exist in
astrophysical and laboratory plasmas. It is found that the dependencies of
wavelength and frequency on wave amplitude at different quantum parameters
for electrons behave differently for compressive and rarefactive
ion-acoustic cnoidal wave cases. The results are useful to understand how
nonlinear wave propagates in quantum plasmas.


\section*{Acknowledgements}

SM acknowledges CNPq (National Council for Scientific and Technological
Development) and TWAS (The World Academy of Sciences) for a CNPq-TWAS
postdoctoral fellowship. FH acknowledges CNPq for financial support.

\end{document}